\documentclass{cs19proc}

\usepackage{kantlipsum}

\editors{G.~A. Feiden}
\publisher{Zenodo}
\conference{The 19th Cambridge Workshop on Cool Stars, Stellar Systems, and the Sun}
\conferencedate{2016}

\title{Planet signatures in the chemical composition of Sun-like stars}
\author{Jorge Mel\'endez$^{1}$, Iv\'an Ram{\'{\i}}rez$^{2}$}

\affiliation{$^{1}$ Departamento de Astronomia, IAG, Universidade de S\~ao Paulo, S\~ao Paulo, Brazil \\
$^{2}$ Tacoma Community College, Washington, USA}

\shorttitle{Planet signatures in Sun-like stars}
\shortauthors{Mel\'endez \& Ram{\'{\i}}rez}

\abs{There are two possible mechanisms to imprint planet signatures in the chemical composition of Sun-like stars: i) dust condensation at the early stages of planet formation, causing a depletion of refractory elements in the gas accreted by the star in the late stages of its formation; ii) planet engulfment, enriching the host star in lithium and refractory elements. We discuss both planet signatures, the influence of galactic chemical evolution, and the importance of
binaries composed of stellar twins as laboratories to verify abundance anomalies imprinted by planets.}

\begin{document}

\maketitle

\section{Chemical signatures imprinted by planet formation}

Metallicity seems to enhance the formation of giant planets, as first suggested by \cite{gon97}.
Subsequent works showed that indeed metal-rich stars have a higher chance of hosting close-in giant planets
\citep[e.g.,][]{fv05,san04}, and that the formation of neptunes and small planets have a much weaker dependence on metallicity
\citep[e.g.,][]{wf15,sch15,buc14,gue10,sou08}.

Besides the effect that the global metallicity of the natal cloud can have on forming different type of planets,
dust condensation, the first stage in the formation of rocky planets and the rocky cores of giant planets, 
will sequester refractory material, causing a non-negligible impact on the composition of the late gas 
accreted onto the star.
This can alter the chemical composition of the convection zone
of Sun-like stars, because the gas that is accreted by the star at the late stages of its formation
would be depleted in refractory elements, making the star slightly
deficient in these elements at the level of just a few 0.01 dex \citep{cha10}.

\begin{figure}
	\centering
	\includegraphics[width=1.00\linewidth]{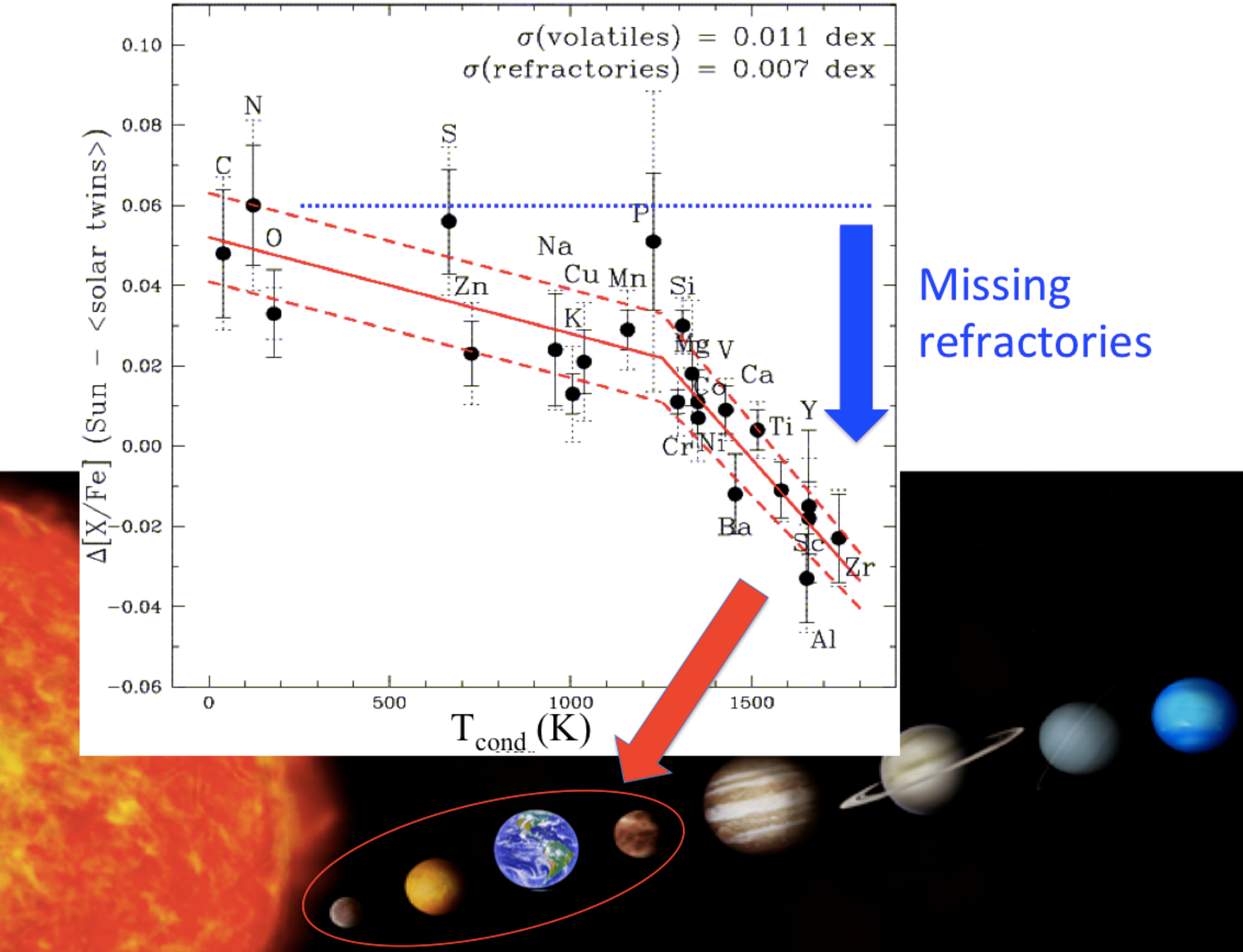}
	\caption{The Sun is slightly depleted in refractory elements  (those with high condensation temperatures) 
	when compared to the average abundance pattern of solar twins. The missing refractories were 
	probably sequestered in the rocky planets and rocky cores of giant planets. 
	Notice that the correlation between abundance anomalies and T$_{\rm cond}$ is quite robust,
	with a probability of 10$^{-9}$ to happen by chance.
        Figure adapted from \cite{mel09}.}
	\label{fig:mel09}
\end{figure}

Remarkably, a comparison of the chemical composition of the Sun to solar twins
by \cite{mel09}, showed that the Sun is deficient in refractory elements (Fig. 1),
an observational finding supported by further works  
\citep{ram09,ram10,gh10,gon10,sch11a,nis15,nis16}.
This fascinating signature of planet formation,
revealed as a strong correlation with condensation temperature (T$_{\rm cond}$)
(Fig. 1), was only unveiled using unprecedentedly precise chemical abundances (0.01 dex),
obtained through a strictly line-by-line differential technique of very similar stars
observed with the same instrument and setup, thus cancelling out many 
uncertainties (e.g., transition probabilities, models atmospheres) and systematic effects that plague abundance analyses
\citep{hin16,bla16,asp05}.

A deficiency of refractory elements has been also observed by \cite{liu16a} 
 in the rocky-planet host {\bf Kepler-10}. As this star has a slightly sub-solar metallicity and
 it is from the thick disk population, the differential analysis was performed using thick disk stars with
 stellar parameters similar to Kepler-10. The abundance pattern of the Galactic thick disk is distinct to
 the thin disk, therefore is mandatory to use reference stars from the same population when looking for 
 the chemical signatures of planets.

One issue regarding the imprint of planet formation signatures in the Sun is that its convective zone
may have been too massive during the lifetime of the disk, diluting the signal to levels that would be very hard to observe.
According to classical stellar evolution models \citep[e.g.,][]{dan94}, it would take about 30 Myr for the Sun's convection zone to shrink to 
about its present-day mass ($\sim$0.02 M$_\odot$), however the lifetime of proto-planetary disks is below 10 Myr \citep[e.g.,][]{mam09,sic09},
albeit there seems to be some dependence with stellar mass, with longer living disks for stars with M $<$ 2 M$_\odot$ \citep{rib15}. 
Note also that \cite{pfa14} have suggested that the observations that have been used to infer fast disk dispersal may be subject to severe selection effects.
A solution to this timescale problem could be the effect of episodic accretion on the stellar structure
\citep{bar10}. The impact of non steady accretion rates would be significantly higher central temperatures,
hence the radiative core develops earlier than in classical models, so that the star can reach a low-mass convective 
envelope in only about 5 Myr, rather than the 30 Myr needed in classical models.

\begin{figure}
	\center
	\includegraphics[width=0.95\linewidth]{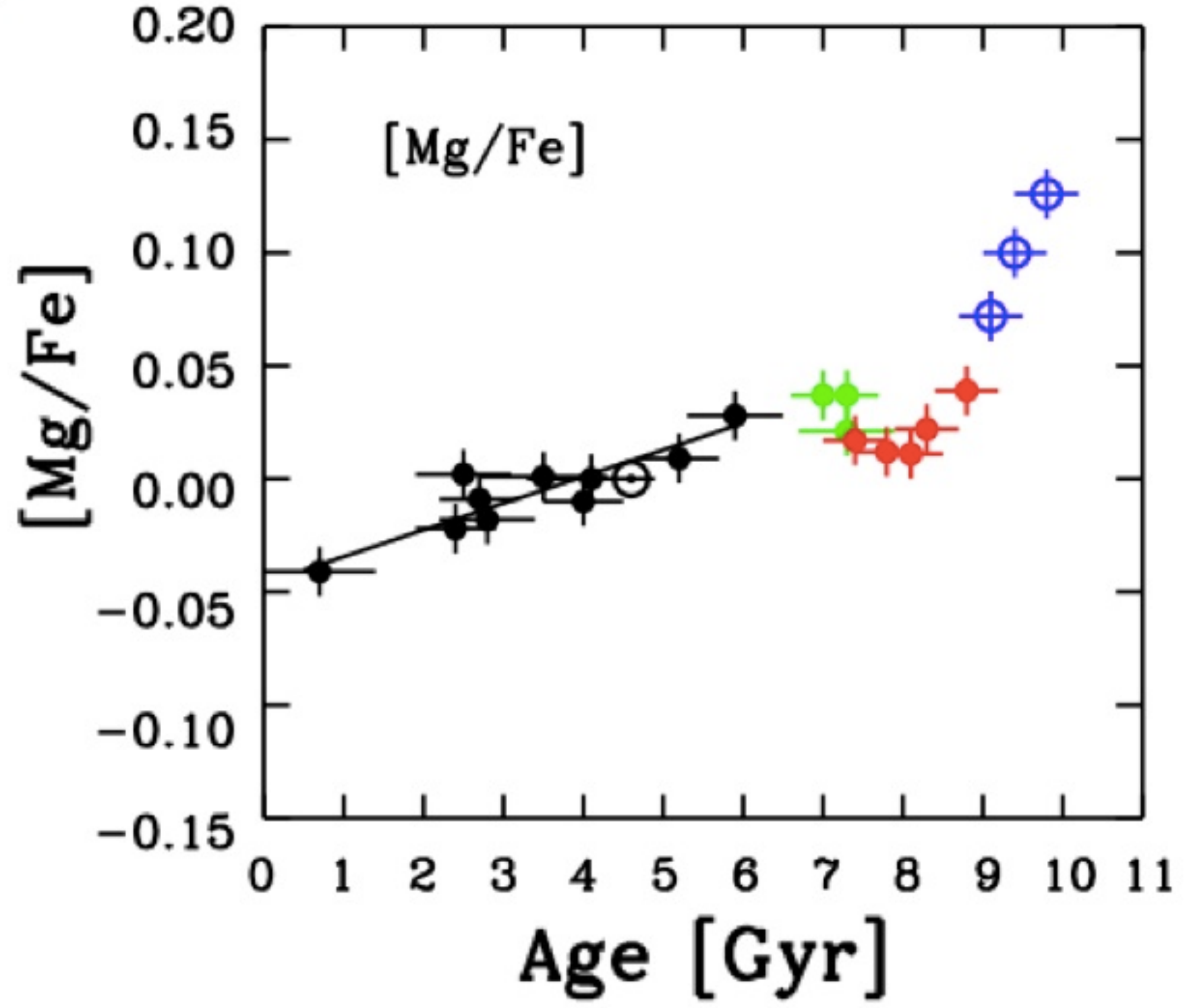}
	\caption{[Mg/Fe] vs. age using solar twins. The three blue circles with the
	highest [Mg/Fe] abundances are old stars likely from the thick disk.
        Figure adapted from \cite{nis16}.}
	\label{fig:nis16}
\end{figure}

\section{Galactic chemical evolution effects}

\begin{figure}
	\centering
	\includegraphics[width=0.95\linewidth]{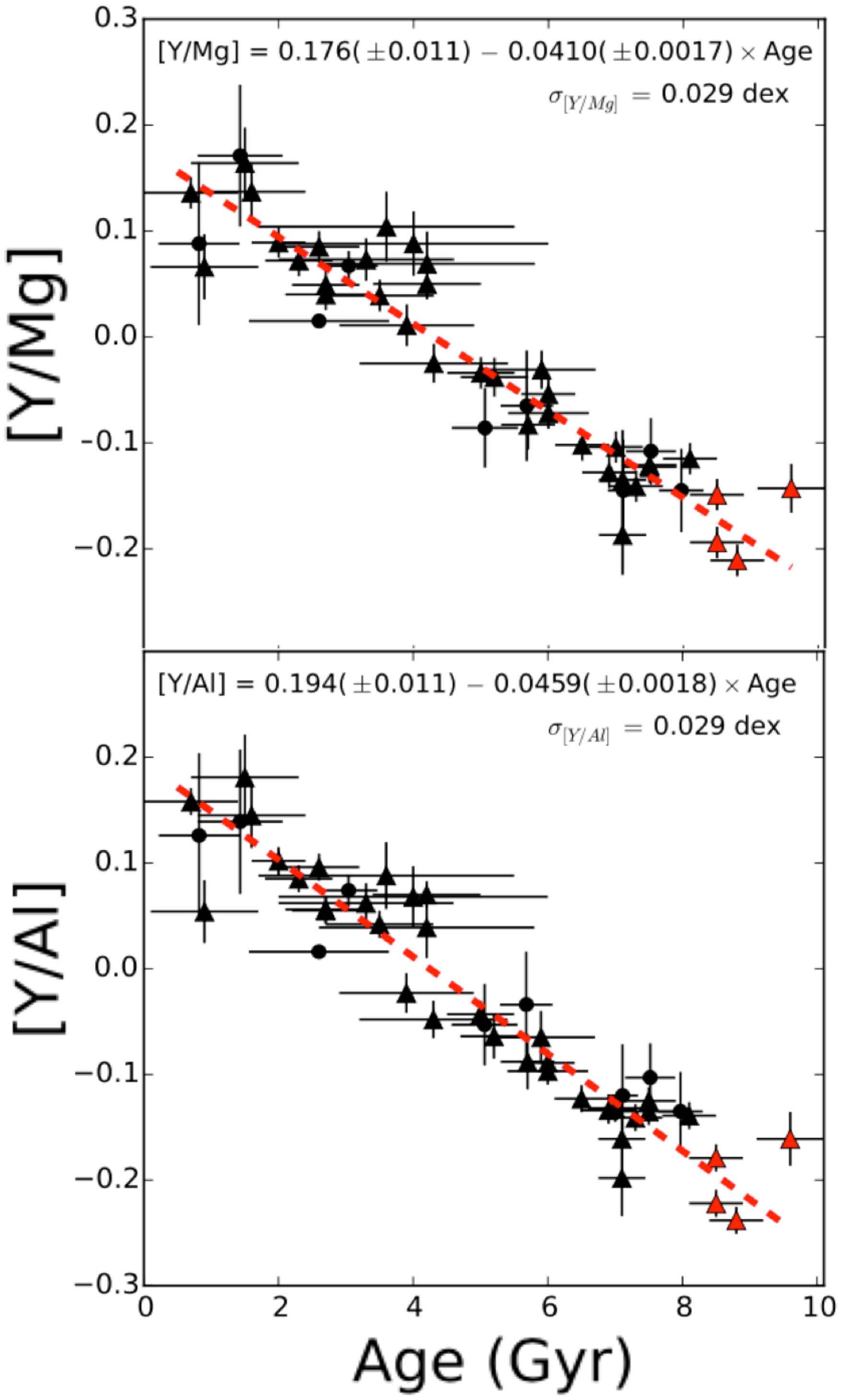}
	\caption{[Y/Mg] and [Y/Al] ratios vs. age based on solar twins.
	The strong dependence of these abundance ratios with age
	arise due to the opposite effects of increasing [Y/Fe] for
	younger stars (due to the increasing contribution by AGB stars) 
	but decreasing [Mg/Fe] and [Al/Fe] ratios.
	The red triangles represent old stars likely from the thick disk population.
        Figure adapted from \cite{spi16b}.}
	\label{fig:spi16}
\end{figure}

\cite{adi14} suggested that there are no signatures of planet formation in the abundance trends of solar analogs,
but instead that the slope with condensation temperature is due mainly to different stellar ages, a 4-$\sigma$ result.
They also tested a correlation with galactocentric distances, but found no significant correlation (at about 1.5-$\sigma$).
\cite{nis15} also tested a correlation between the slope in T$_{\rm cond}$ and age using solar twins, and found a 2.9-$\sigma$ correlation.
In their latest study using late-F and early G dwarfs, \cite{adi16b} confirmed no correlation with galactocentric distance (at the 1.6-$\sigma$ level)
and, unlike their previous 4-$\sigma$ result using a smaller sample, now they found no correlation between 
T$_{\rm cond}$-slope and age.

Albeit age is not the main driver of trends with condensation temperature,
it introduces some variation in the T$_{\rm cond}$-slope due to the scatter introduced by galactic chemical evolution (GCE). 
The variation of individual [X/Fe] ratios with age (e.g., Fig. 2) for elements with Z $\leq 30$, was first observed by \cite{das12}
using solar type stars and more clearly shown by  \cite{nis15} using solar twins. 
In follow-up works, \cite{spi16a,spi16b} found also correlations between
abundance ratios and stellar age.
Interestingly, some abundance ratios that include s-process elements such as yttrium ([Y/Mg] and [Y/Al]),
have a strong dependence with age (Fig. 3),
so that they may be useful for age determination \citep{das12,nis15,nis16,spi16b,tuc16},
at least for near solar composition stars;
but the metallicity-dependence of those relations needs to be established \citep{fel16}.

\begin{figure}
	\centering
	\includegraphics[width=0.99\linewidth]{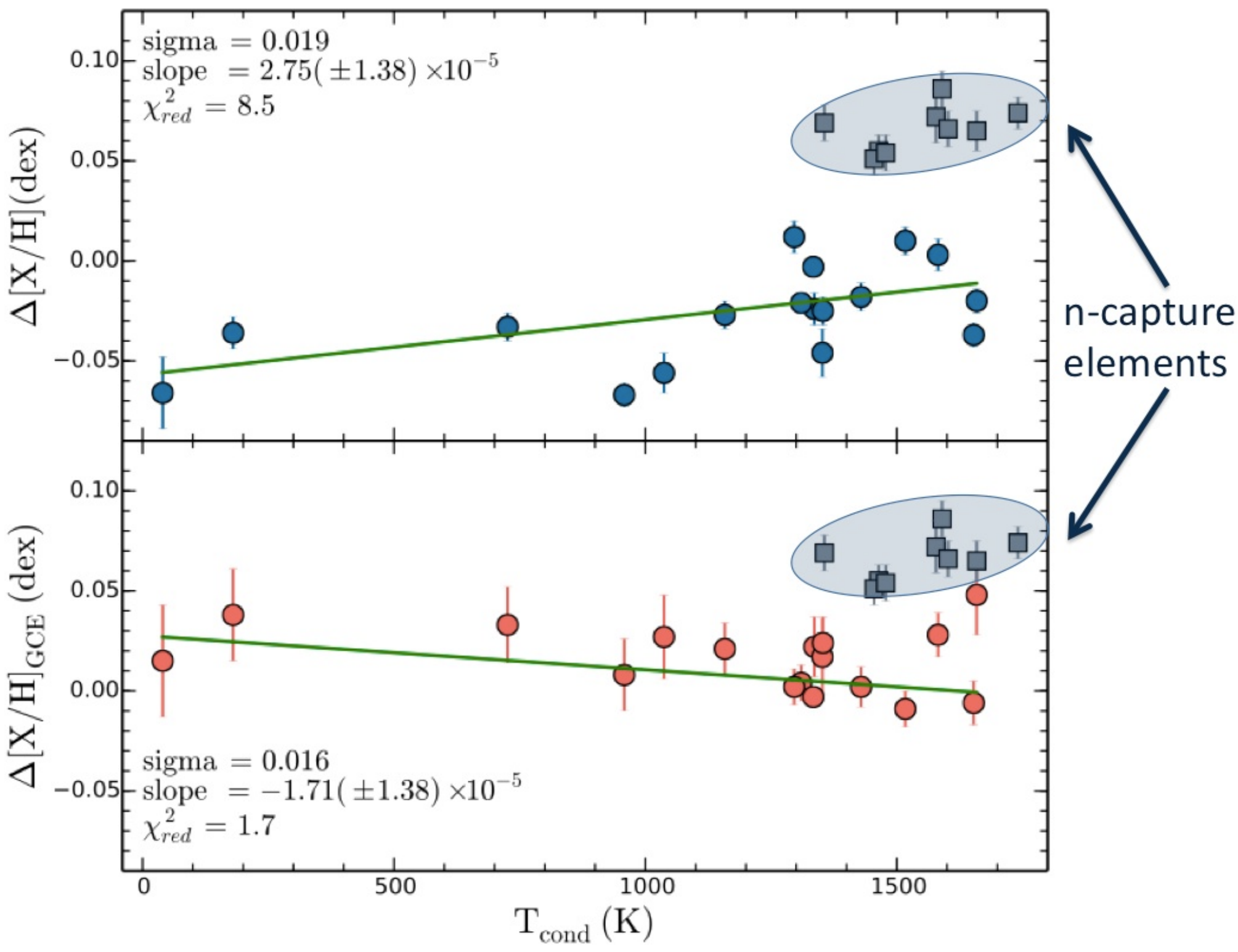}
	\caption{[X/H] ratios vs. T$_{\rm cond}$ in the solar twin HIP 100963. In the upper panel
	the abundance ratios without GCE correction; notice the high reduced $\chi^2$ (8.5).
	In the lower panel the abundance ratios after GCE corrections to the Sun's age,
	with the scatter about the fit now compatible with the abundance errors 
	(including now uncertainties in the GCE corrections) and
	a much lower reduced $\chi^2$ (1.7).
        Figure adapted from \cite{yan16}.}
	\label{fig:yan16}
\end{figure}

\cite{spi16a} suggested that the GCE effects could be
disentangled from the planet signatures by removing the [X/Fe] vs. age trends;
after this correction the Sun is one of the most refractory-poor stars for its age.
Furthermore, modelling the abundances of solar twins using [X/Fe]-age relations and trends with
T$_{\rm cond}$, reduce the scatter of the abundance ratios \citep{spi16b}.
Using a similar procedure to correct for GCE, \cite{nis16} found that the Sun has the most negative slope
with T$_{\rm cond}$ in his sample of solar twins, meaning that the Sun is the most refractory-poor star
even after GCE is taken into account. Thus, the deficiency of refractory elements in the Sun is not
entirely caused by GCE but it could be related to planets.

When comparing Galactic think disk stars with ages differing from the Sun, it is important to
perform GCE corrections. \cite{yan16} showed that the young (2 Gyr) solar twin {\bf HIP 100963}
seems somewhat enhanced in refractories but with a high scatter (0.019 dex) around the fit of [X/H] vs. T$_{\rm cond}$,
more than twice the typical error bar (0.008 dex) in the [X/H] abundance ratios (Fig. 4, upper panel).
After correcting the abundance ratios for GCE to the solar age, the abundance pattern of HIP 100963 seems
actually solar or slightly depleted in refractories (relative to the Sun). Also, the scatter around the fit
is reduced (0.016 dex) and is now compatible with the average error (0.015 dex) that includes the uncertainties in
the GCE corrections (Fig. 4, lower panel).

\section{Binaries as laboratories to study planet signatures}

The study of binaries is key to verify if there are chemical signatures associated with planets.
As both binary components arose from the same natal cloud, it is expected that both stars
present the same abundance pattern. If the components are twins of each other, meaning very close in
stellar parameters (not necessarily solar), precise abundances can be obtained through a careful line-by-line
analysis of high resolution spectra, allowing to verify potential abundance differences that could be due to planets \citep{des06}.

One of the most important binaries to study planet effects is {\bf 16 Cyg}, which consists
of two solar twins, one of them hosting a giant planet \citep{coc97}.
In a fine differential analysis, \cite{law01} found the stars to differ in metallicity
by $\Delta$[Fe/H] = +0.025$\pm$0.009 dex (16 Cyg A - B). A detailed analysis of the system by 
\cite{ram11} showed that indeed 16 Cyg A is more metal-rich than 16 Cyg B, not only in iron but also in 23 other elements 
by about +0.04 dex. Nevertheless, \cite{sch11b} found no abundance differences
between the components. Notice however that considering the error bars, the results by
\cite{sch11b} are consistent with the abundance differences of about 0.04 dex found by \cite{ram11}.


\begin{figure}
	\centering
	\includegraphics[width=0.99\linewidth]{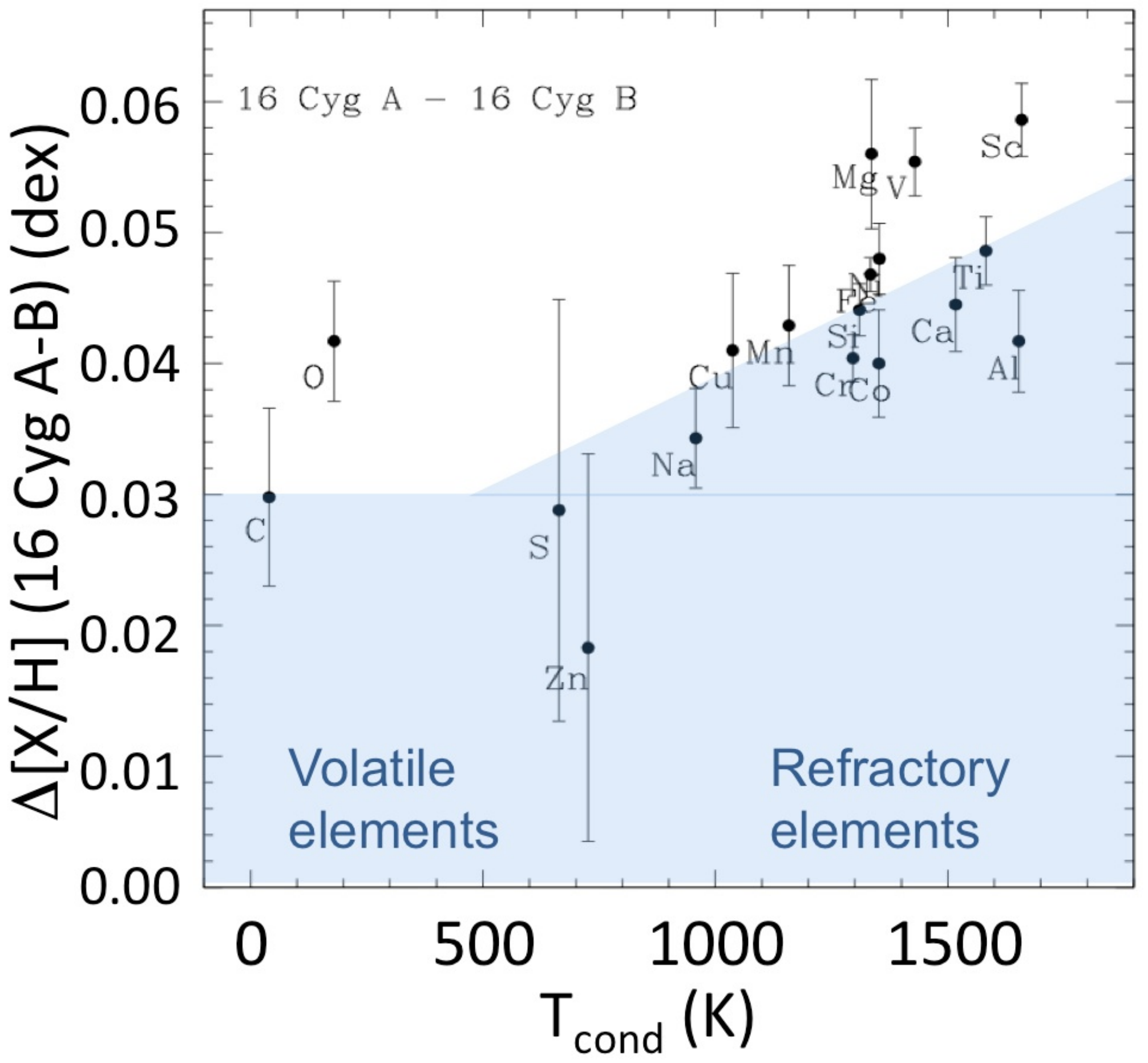}
	\caption{Abundances differences between 16 Cyg A and B vs. condensation temperature,
	according to the most precise analysis of the system by \cite{tuc14}.
	Owing to the extremely high precision,
	most abundance ratios are several sigma above the 0.0 level, showing thus the
	existence of genuine abundance differences between the binary components.  
	Interestingly, the refractory elements seem more enhanced than the volatile elements.
        Figure adapted from \cite{tuc14}.}
	\label{fig:marcelo}
\end{figure}

\begin{figure}
	\centering
	\includegraphics[width=0.99\linewidth]{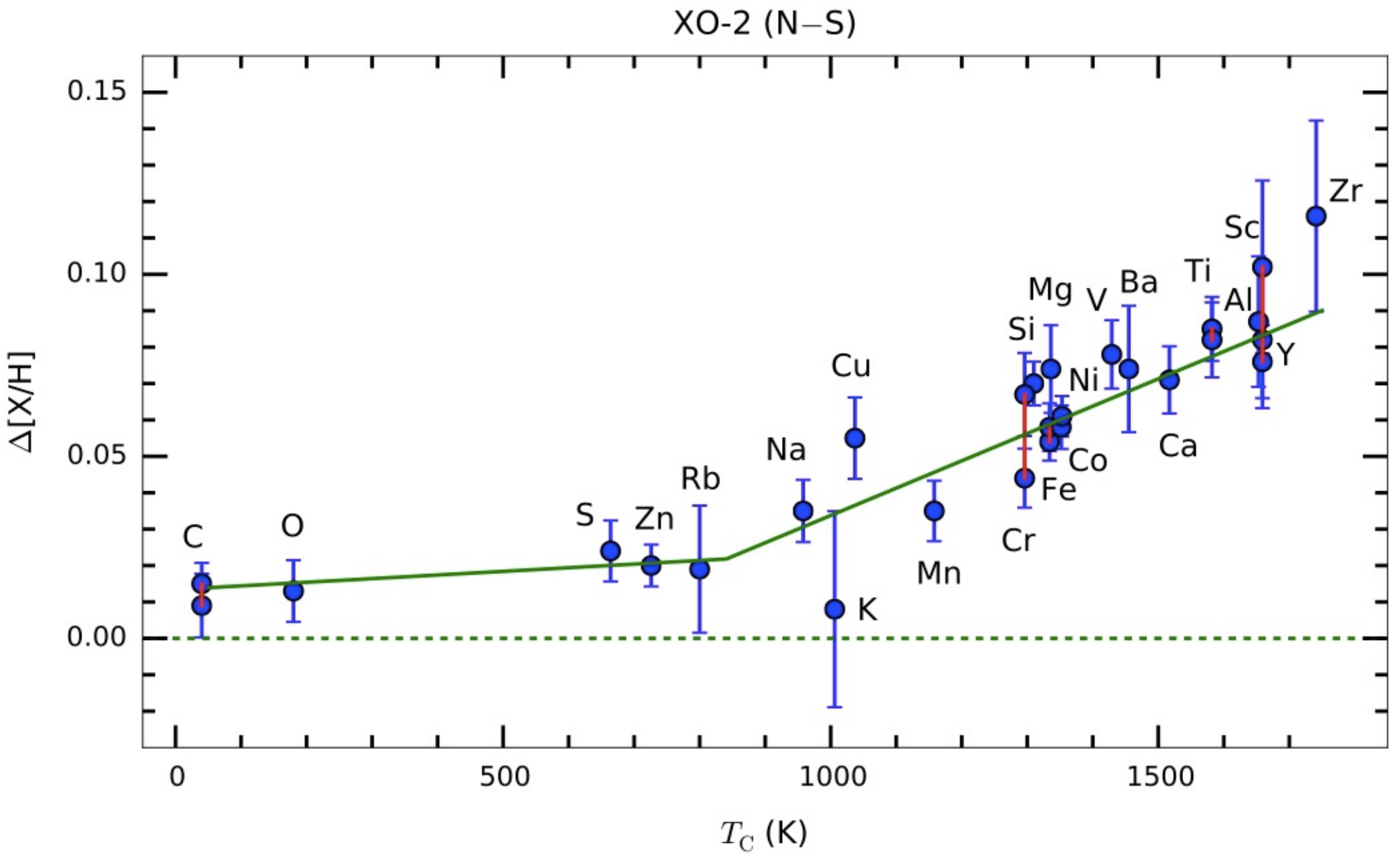}
	\caption{Abundances differences between the planet-hosting stars of the XO-2 binary system.
	Clear abundance differences are found, especially for the refractory elements.
	Figure from \cite{ram15}.}
	\label{fig:xo2}
\end{figure}

Another study using spectra of much higher quality (R = 81 000, S/N = 700) by \cite{tuc14}, 
showed abundance differences between 16 Cyg A and B, considerably above their small error bars
(Fig. 5), suggesting thus that there are genuine abundance differences that could be related
to planets. The most recent study on 16 Cyg \citep{mis16} found also abundance differences
between the components.

Other binaries of stellar twins hosting planets or debris disks 
have been carefully studied for abundance differences. Most of these studies
strongly suggest chemical abundance anomalies due to planets: 
{\bf XO-2N/XO-2S} \citep[][see Fig. 6]{ram15,bia15,tes15};
{\bf $\zeta^{1,2}$ Ret} \citep{saf16,adi16a};
{\bf WASP 94 A/B} \citep{tes16a};
{\bf HD 133131A/B} \citep{tes16b}. 

Albeit abundance anomalies possibly related to planets have been found in 5 binaries so far,
no differences have been found between the two components of the {\bf HAT-P-1} binary
\citep{liu14}. For the systems {\bf HD80606/HD80607} \citep{saf15,mac16} and
{\bf HD20782/HD20781} \citep{mac14}, there are inconclusive results due to relatively high errors in the 
abundances; notice in particular that in the latter system there is a large temperature
difference between the components \citep[$\Delta$ T$_{\rm eff}$ = 465 K,][]{mac14}, making it harder to achieve precise abundances.

\section{Signatures of planet accretion}

Observations of massive neptunes and jupiters in close-in orbits suggest migration from the outer
to inner planetary regions.
It is possible that planet migration events can induce inner planets to move into unstable
orbits and some of them may be engulfed by their host stars, potentially altering their
surface chemical composition \citep{san02}.

Since Sun-like stars deplete lithium as they age \citep[e.g.,][]{car16,and15,mon13,den10,bau10,don09},
it is easy to increase their Li abundance by planet accretion.
Albeit most solar twins analysed by \cite{car16} follow a well-defined Li-age correlation,
they noticed three stars with enhanced Li abundances, among them 16 Cyg A.
Interestingly, this star is also enhanced in volatiles and in refractories (Fig. 5).
This suggests that  the old solar twin 16 Cyg A may not have a planet because it has already 
accreted it.
Note, however, that \cite{dea15} argue that if the effects of fingering convection are taken into consideration, surface pollution by metals from an engulfed planet would be quickly diluted, while lithium would in fact be destroyed by mixing processes, in direct contradiction with the statement made above.

Although many planetary systems show evidences of migration of outer giant
planets to the innermost regions of their systems, our solar system has a
well-defined architecture with small rocky planets in the inner region and giant planets
in the outer region. This contrasting configuration may be due to 
the dynamical barrier imposed by Jupiter, preventing other giant
planets (or large rocky cores or giant planets) from ending up in the inner solar system region \citep{izi15}. 
If so, then the solar twin {\bf HIP 11915} may be an excellent candidate for a solar system twin,
as this star hosts a Jupiter twin (a Jupiter-mass planet in a Jupiter-like orbit) but no other 
giant planets are detected in the inner region \citep{bed15}.
On the other hand, the solar twin {\bf HIP 68468} does not show evidence of a Jupiter analog
but it has an inner (0.7 AU) super-Neptune and potentially a hot super-Earth  \citep{mel16}. 
The lack of a Jupiter analog perhaps allowed the super-Neptune to migrate inwards where it is
observed today. Due to this migration, inner planets may have ended up engulfed by its host star,
as suggested by the enhancement of refractory elements and lithium (Fig. 7)
detected in this star \citep{mel16}.

Another signature of planet accretion is presented by a young star in the {\bf Gamma Velorum} 
open cluster, as revealed by a large enhancement in the refractory
elements, relative to other cluster members \citep{spi15}.
The effects of planet ingestion on this star are discussed in detail by \cite{tog16}.

\begin{figure*}
	\centering
	\includegraphics[width=0.99\linewidth]{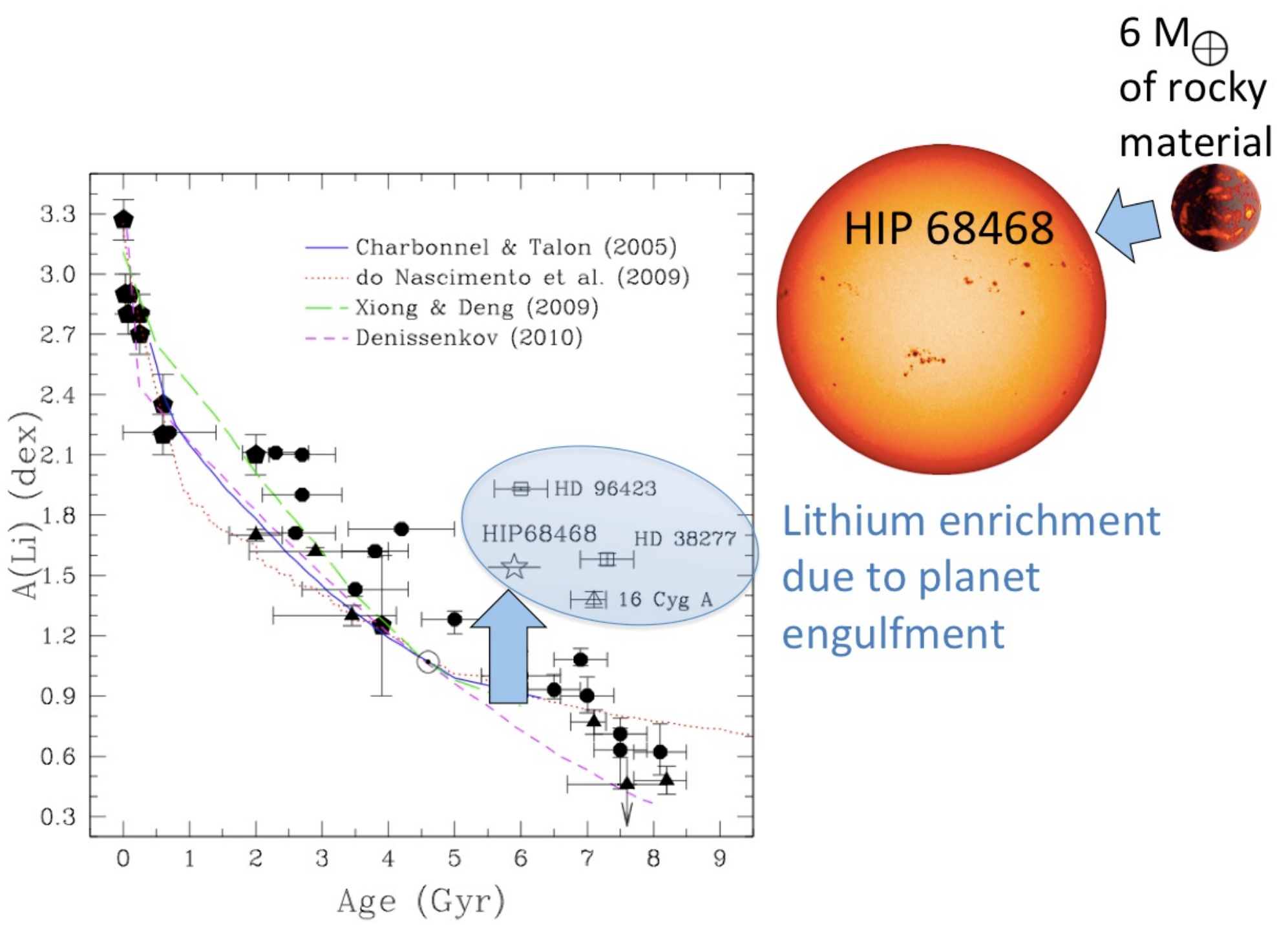}
	\caption{Li-age correlation in solar twins. 
	The enhanced lithium abundance in HIP 68468 may be due to the accretion of 6 Earth masses of rocky material.
	This amount of planet ingestion can also explain the observed enhancement of refractory elements in this solar twin.
	Other solar twins shown in the shaded area may have also engulfed planets \citep{car16}.
        Figure adapted from \cite{mel16}.}
	\label{fig:mel16}
\end{figure*}

\section{Conclusions}

It seems well-established that planet formation and evolution can imprint distinct signatures in the chemical composition
of their host stars. 
This is evident in the analysis of planet-hosting binaries composed of stellar twins, where high precision can be
achieved and abundance anomalies can be related directly to planets.

The abundance ratios are also affected by the chemical evolution of the Galaxy,
but as this variation seems predictable, it could be disentangled from the planetary signature.
After GCE corrections, the Sun is one of the most refractory-poor stars \citep{nis16}, 
strongly suggesting a connection with planet formation.

Besides binaries, open clusters could be used to seek for abundance anomalies due to
planets. Unfortunately not much work have been done at high precision yet,
except for the clusters Hyades \citep{liu16b} and M67 \citep{liu16c}, where small abundance
anomalies have been found but they may not be related to planets.

Further studies connecting exoplanet detection/characterization and high-precision 
chemical abundances in solar twins \citep[e.g.,][]{ram14,bed15,mel16}, 
binaries \citep[e.g.,][]{ram15,tes16a,tes16b,bia15,tuc14} and open clusters \citep[e.g.,][]{liu16c,bru14,bru16,one11},
can reveal unique information about planet formation and evolution. 

\section*{Acknowledgments}
{J. M. would like to acknowledge support from FAPESP (2012/24392-2) and CNPq (Bolsa de Produtividade).}

\bibliographystyle{cs19proc}

\begin{thebibliography}{}
\bibitem[Adibekyan et al.(2014)]{adi14} Adibekyan, V.~Z., Gonz{\'a}lez Hern{\'a}ndez, J.~I., Delgado Mena, E., et al.\ 2014, \aap, 564, L15 

\bibitem[Adibekyan et al.(2016a)]{adi16a} Adibekyan, V., Delgado-Mena, E., Figueira, P., et al.\ 2016a, \aap, 591, A34 

\bibitem[Adibekyan et al.(2016b)]{adi16b} Adibekyan, V., Delgado-Mena, E., Figueira, P., et al.\ 2016b, \aap, 592, A87 

\bibitem[Andr{\'a}ssy \& Spruit(2015)]{and15} Andr{\'a}ssy, R., \& Spruit, H.~C.\ 2015, \aap, 579, A122 

\bibitem[Asplund(2005)]{asp05} Asplund, M.\ 2005, \araa, 43, 481 

\bibitem[Baraffe \& Chabrier(2010)]{bar10} Baraffe, I., \& Chabrier, G.\ 2010, \aap, 521, A44 

\bibitem[Baumann et al.(2010)]{bau10} Baumann, P., Ram{\'{\i}}rez, I., Mel{\'e}ndez, J., Asplund, M., \& Lind, K.\ 2010, \aap, 519, A87 

\bibitem[Bedell et al.(2015)]{bed15} Bedell, M., Mel{\'e}ndez, J., Bean, J.~L., et al.\ 2015, \aap, 581, A34 

\bibitem[Biazzo et al.(2015)]{bia15} Biazzo, K., Gratton, R., Desidera, S., et al.\ 2015, \aap, 583, A135 

\bibitem[Blanco-Cuaresma et al.(2016)]{bla16} Blanco-Cuaresma, S., Nordlander, T., Heiter, U., et al.\ 2016,  The 19th Cambridge Workshop on Cool Stars, Stellar Systems, and the Sun, arXiv:1609.08092 

\bibitem[Brucalassi et al.(2014)]{bru14} Brucalassi, A., Pasquini, L., Saglia, R., et al.\ 2014, \aap, 561, L9 

\bibitem[Brucalassi et al.(2016)]{bru16} Brucalassi, A., Pasquini, L., Saglia, R., et al.\ 2016, \aap, 592, L1 

\bibitem[Buchhave et al.(2014)]{buc14} Buchhave, L.~A., Bizzarro, M., Latham, D.~W., et al.\ 2014, \nat, 509, 593 

\bibitem[Carlos et al.(2016)]{car16}  Carlos, M., Nissen, P.~E., \& Mel{\'e}ndez, J.\ 2016, \aap, 587, A100 

\bibitem[Chambers(2010)]{cha10} Chambers, J.~E.\ 2010, \apj, 724, 92 

\bibitem[Cochran et al.(1997)]{coc97} Cochran, W.~D., Hatzes, A.~P., Butler, R.~P., \& Marcy, G.~W.\ 1997, \apj, 483, 457 

\bibitem[D'Antona \& Mazzitelli(1994)]{dan94} D'Antona, F., \& Mazzitelli, I.\ 1994, \apjs, 90, 467 

\bibitem[da Silva et al.(2012)]{das12} da Silva, R., Porto de Mello, G.~F., Milone, A.~C., et al.\ 2012, \aap, 542, A84 

\bibitem[Deal et al.(2015)]{dea15} Deal, M., Richard, O., \& Vauclair, S.\ 2015, \aap, 584, A105 

\bibitem[Denissenkov(2010)]{den10} Denissenkov, P.~A.\ 2010, \apj, 719, 28 

\bibitem[Desidera et al.(2006)]{des06} Desidera, S., Gratton, R.G., Lucatello, S., \& Claudi, R.U. \ 2006, \aap, 454, 581

\bibitem[do Nascimento et al.(2009)]{don09} Do Nascimento, J.~D., Jr., Castro, M., Mel{\'e}ndez, J., Bazot, M., Th{\'e}ado, S., Porto de Mello, G.~F., \& de Medeiros, J.~R.\ 2009, \aap, 501, 687 

\bibitem[Feltzing et al.(2016)]{fel16} Feltzing, S., Howes, L.~M., McMillan, P.~J., \& Stonkute, E.\ 2016, arXiv:1610.03852 

\bibitem[Fischer \& Valenti(2005)]{fv05} Fischer, D.~A., \& Valenti, J.\ 2005, \apj, 622, 1102 

\bibitem[Ghezzi et al.(2010)]{gue10} Ghezzi, L., Cunha, K., Smith, V.~V., et al.\ 2010, \apj, 720, 1290 

\bibitem[Gonzalez(1997)]{gon97} Gonzalez, G.\ 1997, \mnras, 285, 403 

\bibitem[Gonzalez et al.(2010)]{gon10} Gonzalez, G., Carlson, M.~K., \& Tobin, R.~W.\ 2010, \mnras, 407, 314 

\bibitem[Gonz{\'a}lez Hern{\'a}ndez et al.(2010)]{gh10} Gonz{\'a}lez Hern{\'a}ndez, J.~I., Israelian, G., Santos, N.~C., Sousa, S., Delgado-Mena, E., Neves, V., \& Udry, S.\ 2010, \apj, 720, 1592 

\bibitem[Hinkel et al.(2016)]{hin16} Hinkel, N.~R., Young, P.~A., Pagano, M.~D., et al.\ 2016, \apjs, 226, 4 

\bibitem[Izidoro et al.(2015)]{izi15} Izidoro, A., Raymond, S.~N., Morbidelli, A., Hersant, F., \& Pierens, A.\ 2015, \apjl, 800, L22 

\bibitem[Laws \& Gonzalez(2001)]{law01} Laws, C., \& Gonzalez, G.\ 2001, \apj, 553, 405 

\bibitem[Liu et al.(2014)]{liu14} Liu, F., Asplund, M., Ram{\'{\i}}rez, I., Yong, D., \& Mel{\'e}ndez, J.\ 2014, \mnras, 442, L51 

\bibitem[Liu et al.(2016a)]{liu16a} Liu, F., Yong, D., Asplund, M., et al.\ 2016a, \mnras, 456, 2636 

\bibitem[Liu et al.(2016b)]{liu16b} Liu, F., Yong, D., Asplund, M., Ram{\'{\i}}rez, I., \& Mel{\'e}ndez, J.\ 2016b, \mnras, 457, 3934 

\bibitem[Liu et al.(2016c)]{liu16c} Liu, F., Asplund, M., Yong, D., et al.\ 2016c, \mnras, 463, 696 

\bibitem[Mack et al.(2014)]{mac14} Mack, C.~E., III, Schuler, S.~C., Stassun, K.~G., \& Norris, J.\ 2014, \apj, 787, 98 

\bibitem[Mack et al.(2016)]{mac16} Mack, C.~E., III, Stassun, K.~G., Schuler, S.~C., Hebb, L., \& Pepper, J.~A.\ 2016, \apj, 818, 54 

\bibitem[Mamajek(2009)]{mam09} Mamajek, E.~E.\ 2009, American Institute of Physics Conference Series, 1158, 3 

\bibitem[Mel{\'e}ndez et al.(2009)]{mel09} Mel{\'e}ndez, J., Asplund, M., Gustafsson, B., \& Yong, D.\ 2009, \apjl, 704, L66

\bibitem[Mel\'endez et al.(2016)]{mel16} Mel\'endez, J., Bedell, M., Bean, J.~L., et al.\ 2016, \aap, in press, arXiv:1610.09067 

\bibitem[Mishenina et al.(2016)]{mis16} Mishenina, T., Kovtyukh, V., Soubiran, C., \& Adibekyan, V.~Z.\ 2016, \mnras, 462, 1563 

\bibitem[Monroe et al.(2013)]{mon13} Monroe, T.~R., Mel{\'e}ndez, J., Ram{\'{\i}}rez, I., et al.\ 2013, \apjl, 774, L32 

\bibitem[Nissen(2015)]{nis15} Nissen, P.~E.\ 2015, \aap, 579, A52 

\bibitem[Nissen(2016)]{nis16} Nissen, P.~E.\ 2016, \aap, 593, A65 

\bibitem[{\"O}nehag et al.(2011)]{one11} {\"O}nehag, A., Korn, A., Gustafsson, B., Stempels, E., \& Vandenberg, D.~A.\ 2011, \aap, 528, A85

\bibitem[Pfalzner et al.(2014)]{pfa14} Pfalzner, S., Steinhausen, M., \& Menten, K.\ 2014, \apjl, 793, L34 

\bibitem[Ram{\'{\i}}rez et al.(2009)]{ram09} Ram{\'{\i}}rez, I., Mel{\'e}ndez, J., \& Asplund, M.\ 2009, \aap, 508, L17 

\bibitem[Ram{\'{\i}}rez et al.(2010)]{ram10} Ram{\'{\i}}rez, I., Asplund, M., Baumann, P., Mel{\'e}ndez, J., \& Bensby, T.\ 2010, \aap, 521, A33

\bibitem[Ram{\'{\i}}rez et al.(2011)]{ram11} Ram{\'{\i}}rez, I., Mel{\'e}ndez, J., Cornejo, D., Roederer, I.~U., \& Fish, J.~R.\ 2011, \apj, 740, 76 

\bibitem[Ram{\'{\i}}rez et al.(2014)]{ram14} Ram{\'{\i}}rez, I., Mel{\'e}ndez, J., Bean, J., et al.\ 2014, \aap, 572, A48 

\bibitem[Ram{\'{\i}}rez et al.(2015)]{ram15} Ram{\'{\i}}rez, I., Khanal, S., Aleo, P., et al.\ 2015, \apj, 808, 13 

\bibitem[Ribas et al.(2015)]{rib15} Ribas, {\'A}., Bouy, H., \& Mer{\'{\i}}n, B.\ 2015, \aap, 576, A52 

\bibitem[Saffe et al.(2015)]{saf15} Saffe, C., Flores, M., \& Buccino, A.\ 2015, \aap, 582, A17 

\bibitem[Saffe et al.(2016)]{saf16} Saffe, C., Flores, M., Jaque Arancibia, M., Buccino, A., \& Jofr{\'e}, E.\ 2016, \aap, 588, A81 

\bibitem[Sandquist et al.(2002)]{san02} Sandquist, E.~L., Dokter, J.~J., Lin, D.~N.~C., \& Mardling, R.~A.\ 2002, \apj, 572, 1012 

\bibitem[Santos et al.(2004)]{san04} Santos, N.~C., Israelian, G., \& Mayor, M.\ 2004, \aap, 415, 1153 

\bibitem[Schuler et al.(2011a)]{sch11a} Schuler, S.~C., Flateau, D., Cunha, K., King, J.~R., Ghezzi, L., \& Smith, V.~V.\ 2011a, \apj, 732, 55 

\bibitem[Schuler et al.(2011b)]{sch11b} Schuler, S.~C., Cunha, K., Smith, V.~V., et al.\ 2011b, \apjl, 737, L32 

\bibitem[Schuler et al.(2015)]{sch15} Schuler, S.~C., Vaz, Z.~A., Katime Santrich, O.~J., et al.\ 2015, \apj, 815, 5 

\bibitem[Sicilia-Aguilar et al.(2009)]{sic09} Sicilia-Aguilar, A., Bouwman, J., Juh{\'a}sz, A., et al.\ 2009, \apj, 701, 1188 

\bibitem[Sousa et al.(2008)]{sou08} Sousa, S.~G., Santos, N.~C., Mayor, M., et al.\ 2008, \aap, 487, 373 

\bibitem[Spina et al.(2015)]{spi15} Spina, L., Palla, F., Randich, S., et al.\ 2015, \aap, 582, L6 

\bibitem[Spina et al.(2016a)]{spi16a} Spina, L., Mel{\'e}ndez, J., \& Ram{\'{\i}}rez, I.\ 2016a, \aap, 585, A152 

\bibitem[Spina et al.(2016b)]{spi16b} Spina, L., Mel{\'e}ndez, J., Karakas, A.~I., et al.\ 2016b, \aap, 593, A125 

\bibitem[Teske et al.(2015)]{tes15} Teske, J.~K., Ghezzi, L., Cunha, K., et al.\ 2015, \apjl, 801, L10 

\bibitem[Teske et al.(2016a)]{tes16a} Teske, J.~K., Khanal, S., \& Ram{\'{\i}}rez, I.\ 2016a, \apj, 819, 19 

\bibitem[Teske et al.(2016b)]{tes16b} Teske, J.~K., Shectman, S.~A., Vogt, S.~S., et al.\ 2016b, \aj, in press, arXiv:1608.06216 

\bibitem[Tognelli et al.(2016)]{tog16} Tognelli, E., Prada Moroni, P.~G., \& Degl'Innocenti, S.\ 2016, \mnras, 460, 3888 

\bibitem[Tucci Maia et al.(2014)]{tuc14} Tucci Maia, M., Mel{\'e}ndez, J., \& Ram{\'{\i}}rez, I.\ 2014, \apjl, 790, LL25

\bibitem[Tucci Maia et al.(2016)]{tuc16} Tucci Maia, M., Ram{\'{\i}}rez, I., Mel{\'e}ndez, J., et al.\ 2016, \aap, 590, A32 

\bibitem[Wang \& Fischer(2015)]{wf15} Wang, J., \& Fischer, D.~A.\ 2015, \aj, 149, 14 

\bibitem[Yana Galarza et al.(2016)]{yan16} Yana Galarza, J., Mel{\'e}ndez, J., Ram{\'{\i}}rez, I., et al.\ 2016, \aap, 589, A17 

\end{thebibliography}

\end{document}